\documentclass[aps,tightenlines,preprint]{revtex4-1}
\usepackage{graphics,psfrag,amsmath}
\usepackage{epic,eepic}
\usepackage{dcolumn}
\usepackage{bm}
\usepackage{lmodern}
\usepackage{color} 
\usepackage{epsfig}
\usepackage{latexsym}
\usepackage{pstricks}
\newcommand{\be}{\begin{equation}}
\newcommand{\ee}{\end{equation}}
\newcommand{\bear}{\begin{eqnarray}}
\newcommand{\eear}{\end{eqnarray}}
\begin{document}

\title{Bending of light  in a Coulomb Gas}
\author{Taekoon Lee}
\email{tlee@kunsan.ac.kr}
\affiliation{Department of 
Physics, Kunsan National University, Kunsan 54150, Korea}


\begin{abstract}
Photons traveling in a background electromagnetic field may bend via the
vacuum polarization effect with the background field. The bending 
in a Coulomb field by a heavy nucleus is small even 
at a large atomic number, rendering it difficult to detect 
experimentally. As an amplifying mechanism of the effect we consider 
the bending of light traveling in a chamber of Coulomb gas. The 
Gaussian nature of the  bending  in the  gas increases the 
total bending angle in proportion to the square root of the
photon travel distance. The enhancement can be 
orders of magnitude over the 
bending by a single nucleus at a small impact 
parameter, which may help experimental 
observation of  the Coulombic 
bending.
\end{abstract}  \pacs{}  

\maketitle  
The vacuum polarization effect of the quantum 
electrodynamics  renders 
a photon traveling in a background electromagnetic field bend. For a 
photon moving in a 
slowly varying background field,  with energy smaller than the 
electron rest 
mass, the bending may be described by a 
low-energy effective Lagrangian that encodes the vacuum polarization 
effect. At the leading order of the fine structure constant the 
polarization effect is given by a box diagram with four external photon 
lines, which gives rise to the nonlinear interaction of
Euler-Heisenberg \cite{euler,schwinger}:
\bear
{\cal L}= -\frac{1}{4} F_{\mu\nu}F^{\mu\nu} +
\frac{\alpha^2\hbar^3}{90m_{e}^4c^{5}}
\left[( F_{\mu\nu}F^{\mu\nu})^2+\frac{7}{4}
( F_{\mu\nu}\tilde F^{\mu\nu})^2\right]\,,
\label{lagrangian}
\eear  
where $\hbar, c, \alpha$, and $m_{e}$  are the Planck constant, the speed of light, the
fine structure constant, and the electron mass, respectively,
 and  $\tilde{F}_{\mu\nu}$ denotes the dual of the 
field strength tensor $F_{\mu\nu}$. 

A linearization of the Euler-Heisenberg 
interaction in a slowly varying background field yields a photon 
dispersion relation in which the background field is encapsulated in 
an index of refraction $n$
\cite{bialynicka,adler,heyl,boer,novello}:
\bear
n=1+\frac{a\alpha^2\hbar^3}{45m_{e}^4c^5}\left[\bm{E}^{2}+
\bm{B}^{2}-(\bm{\hat{k}}\cdot\bm{E})^{2}
-(\bm{\hat{k}}\cdot\bm{B})^{2}-
2\bm{\hat{k}}\cdot(\bm{E}\times\bm{B})\right]\,,
\eear
where $\bm{\hat{k}}$ is the unit vector 
in the direction of  photon propagation,
 and $a$
 is the birefringence constant 
that is either 8 or 14, depending on the photon polarization.

 Thus a photon moving in a background field 
behaves as if it is traveling in a dielectric medium with 
a refractive index that depends on the background field strength, and 
consequently the photon bends when the field strength is nonuniform. 

This light bending has been studied in relation to astronomical objects 
with strong electromagnet fields, such as  
magnetars or black holes \cite{denisov,lorenci,lee2}. On the opposite 
scale, at a microscopic level a particularly interesting problem
 is the 
bending in a Coulomb field. Because the field-dependent index of refraction 
becomes larger at a stronger field, the incoming photon bends toward
the charge, in a fashion reminiscent of
the gravitational bending in general relativity. 

 The bending angle 
can be easily calculated in geometrical optics.
 For a photon with the 
impact parameter $b$ in the Coulomb field by a nucleus of charge $Ze$, 
 it is given by \cite{lee1}

\bear{} 
\theta(b)=\frac{a Z^{2}\alpha^3}{160} 
\left(\frac{\lambdabar_e}{b}\right)^4\,, 
\label{bendingangle} 
\eear{}
 where $\alpha$ is the fine structure constant, and
 $\lambdabar_{e}=\hbar/m_{e}c$ is 
the reduced electron Compton length.

The impact 
parameter in Eq. (\ref{bendingangle}) cannot be arbitrarily small, 
putting a limit on the size of the bending angle. Requiring that the 
Euler-Heisenberg interaction be a small perturbation to the Maxwell 
theory places a constraint on the field strength \cite{bialynicka}: 
\bear{} 
\frac{2a\alpha^{2}\hbar^{3}}{45m_{e}^{4}c^{5}}
 |F_{\mu\nu}|^{2} \ll 1\,, 
\eear{}
 where $F_{\mu\nu}$ denotes the background  field strength. 
For the Coulomb field  
 \bear{} E(r)=\frac{Ze}{4\pi 
r^{2}} 
\label{coulombfield}
\eear{}
 the constraint requires  the radius to satisfy
  \bear{} r\gg 
\lambdabar_{e}\left(\frac{aZ^{2}\alpha^{3}}{90}\right)^{\frac{1}{4}}\,, 
\label{const1}
\eear{} 
where the fine structure constant is given by 
$\alpha=e^{2}/4\pi\hbar c$.
 Even for a large $Z$ the radius satisfying the constraint can be 
fairly small. For instance, at $Z=100$ 
\bear{} r\gg 0.14 \lambdabar_{e}\,. 
\eear{} 
Also the requirement that the background field be slowly 
varying demands the following \cite{bialynicka}:
\[ 
|\partial_{\lambda}F_{\mu\nu}|\ll \frac{m_{e}c}{\hbar}
|F_{\mu\nu}|\,,\] 
which, for the 
Coulomb field (\ref{coulombfield}), is satisfied when
\bear{} r\gg \lambdabar_{e}\,.{} 
\label{const2} 
\eear{}

We also note that at a very small impact parameter 
where the electric field becomes strong the corrections
to the Euler-Heisenberg interaction can be significant. The
corrections arise from the box diagrams with more than four
external photon lines. 
A simple  dimensional analysis shows that these give rise to 
 effective interactions
in powers of 
\bear{}
\frac{\alpha\hbar^{3}}{m_{e}^{4}c^{5}}
 |F_{\mu\nu}|^{2}\,,
\eear{}
relative to the Euler-Heisenberg term.
For these corrections to be small on
 the Coulomb field (\ref{coulombfield}) the radius must satisfy: 
\bear{}
r\gg \sqrt{Z\alpha}\lambdabar_{e}\,.
\label{const3}
\eear{}
The combined constraints of Eq. (\ref{const1}), (\ref{const2}), 
and (\ref{const3}) on the radius put a
 limit on the impact parameter
in the bending angle  (\ref{bendingangle}).
For heavy nuclei with $Z\alpha\sim 1,$ it requires
\bear{}
b\gg \lambdabar_{e}\,.{}
\label{b-constraint}
\eear{}
  The bending angle under this constraint
is quite small, even at a large $Z$ and  a small impact parameter. For 
instance, for $Z=100$ and $b=10\lambdabar_{e},$ the 
bending is $34$ nano rad.

Because of the smallness of the 
 effect detecting the bending experimentally may be challenging. It may
thus be interesting to study an amplifying mechanism  for the effect.
 As such a  mechanism we consider in this paper a photon (in a 
 collimated beam) traveling in a 
chamber of Coulomb gas that comprises heavy nuclei.

As the photon travels in the chamber it will     
bend off each nuclei in the gas, and because the bendings are random in 
the impact parameter as well as in the azimuthal angle to the beam 
axis, the distribution of the total bending angles will be Gaussian and 
the   Root-Mean-Square (RMS)  angle  be 
proportional to the square root of the number of nuclei in the gas. 
An experimental consequence of this Gaussian bending will 
be a broadening of the photon beam  as it 
travels through the gas.

 Though securing the 
Coulomb gas is beyond the scope of this paper, we could imagine obtaining it
 by blowing 
off the valence electrons of the heavy atoms 
through an illumination of x rays or $\gamma$ rays,
 perhaps as well with the help of an 
electric field applied to the chamber to separate the electrons from 
the nuclei. 
Further, it may not be necessary to separate the electrons
from the ionized nuclei, 
because the bendings off the electrons would be ignorable
 compared to those off the
heavy nuclei. In this case 
 ionization of the heavy atoms alone would suffice for the purpose. 
 Furthermore, even ionization may not be 
necessary with high energy photons (x ray, or $\gamma$ ray), as in 
experiments for the Delbr{\"u}ck scattering \cite{schumacher},
 because the Coulombic 
interactions of the photons with the nuclei would
 occur deep inside the atoms near the nuclei.
In this case the Rayleigh 
scattering off the electrons  must be accurately subtracted.

Now to compute the RMS angle we consider a cylindrical chamber 
of length $L$ and radius $R$ filled with a Coulomb gas 
and assume the photon travels 
along the axis of the 
chamber. Although we start with this particular form of
chamber the final result 
will be independent of the chamber geometry.

To be specific, for a photon with incoming velocity $\vec{c}=c\hat{n}$, 
where $\hat{n}$ denotes the unit vector in the beam direction, the exit
velocity $\vec{v}_{e}$ off the chamber can be written as
\bear{}
\vec{v}_{e}\approx \vec{c}+ \sum_{i=1}^{N} \vec{v}_{\perp}^{i}\,,
\eear{}
where $N$ is the number of nuclei in the Coulomb gas, and 
$\vec{v}_{\perp}^{i}$ denotes the perpendicular component of the beam 
axis of the deflected velocity vector off 
the ${\it i}$-th nucleus.
The total RMS angle $\bar{\Theta}$ is then given by
\bear{}
\bar{\Theta}\equiv\sqrt{\left<\frac{(\vec{v}_{e}-\vec{c})^{2}}{c^{2}}\right>}=
\sqrt{\left<\left(\sum_{i=1}^{N}\frac{\vec{v}_{\perp}^{i}}{c}\right)^{2}\right>}
=\sqrt{\sum_{i=1}^{N}\left<\left(\frac{\vec{v}_{\perp}^{i}}{c}\right)^{2}\right>}
=\sqrt{N}\bar{\theta} \,,
\eear  
where 
\bear{}
\bar{\theta}\equiv\sqrt{\left<\left(\frac{\vec{v}_{\perp}^{i}}{c}\right)^{2}\right>}
\eear{}
is the RMS angle of the bending off a single nucleus.
Noticing that \[\left|\frac{\vec{v}_{\perp}^{i}}{c}\right|\]
is nothing but 
the bending angle $\theta(b)$ in Eq. (\ref{bendingangle}),
 the $\bar{\theta}(b)$ can be computed by
\bear{}
\bar{\theta}=\sqrt{\left<\theta^{2}\right>}\,,
\eear{}
where
\bear{}
\left<\theta^{2}\right>=\int_{\Lambda}^{R}
\langle\theta(b)^{2}\rangle_{\text{spin}} 
P(b)db\,.
\eear{}
Here the averaging over spins is over the photon polarizations, which
applies to the birefringence constant, $\Lambda$ is 
 the lower cutoff in the impact parameter, and $P(b)$ 
denotes the probability density for a particular
 nucleus to fall at the impact parameter 
$b$ with the photon.  The cutoff $\Lambda$ should be subject to the
bound on the impact parameter (\ref{b-constraint}).
Because in a uniformly distributed gas
$P(b)$ should be 
proportional to $b$, the  normalized density is given by
\bear{}
P(b)=\frac{2}{R^{2}}b \,,
\eear{}
which satisfies
\bear{}
\int_{0}^{R}P(b)db=1\,.
\eear{}
Then we get
\bear{}
\bar{\theta}=\frac{\bar{a} 
Z^{2}\alpha^3}{160\sqrt{3}}\frac{\lambdabar_{e}^4}{R\Lambda^{3}} \,,
\eear{}
where 
\bear{}
\bar{a}=\sqrt{(8^{2}+14^{2})/2}=\sqrt{130}
\eear{}
is the RMS of the birefringence constant.
The total RMS angle is then given by
\bear{}
\bar{\Theta}=\frac{
Z^{2}\alpha^3}{160}\sqrt{\frac{130\pi}{3}}
\sqrt{\frac{L}{L_{0}}}
\left(\frac{\lambdabar_{e}}{\Lambda}\right)^{4}\,,{}
\eear{}
where $L_{0}=1/\rho_{\rm N}\Lambda^{2}$, 
with $\rho_{\rm N}=N/V$ denoting the 
number density of the gas in volume $V=\pi R^{2} L$.

Now to estimate the amplification effect of this result we need to
express the number density in terms of temperature and pressure using
the equation of state of the Coulomb gas, which at high density
is not known. However, this problem can be avoided if we assume that 
 the Coulomb gas was obtained in the manner described before, 
 by stripping the electrons off 
 neutral heavy atoms. 
 Then the number density of the Coulomb gas is identical to that
 of the atomic gas, for which we may assume the ideal gas 
law $\rho_{\rm N}=P/k_{\rm B} T$, with $T$ and $P$ 
denoting  the temperature and the pressure of the atomic gas and $k_{\rm B}$
 being the Boltzmann constant. 
 We then have
\bear{}
\bar{\Theta}=\frac{
Z^{2}\alpha^3}{160}\sqrt{\frac{130\pi}{3}}
\sqrt{\frac{L}{L_{0}(P,T,\Lambda)}}
\left(\frac{\lambdabar_{e}}{\Lambda}\right)^{4}\,,{}
\label{result}
\eear{}
where
\[
L_{0}(P,T,\Lambda)=\frac{k_{\rm B}T}{P\Lambda^{2}}\,.
\]

 The result shows that at a given temperature and pressure 
the bending angle increases in 
proportion to the square root of the photon travel distance.
 As asserted, the result  is independent of the geometry of the 
chamber, as there is no geometry-dependent parameter except for $L$, 
which, however, being the distance of the 
photon traveled,  is not particular to the
geometry.

To see the amplifying effect at some 
readily available  parameter values we write Eq. (\ref{result}) as
\bear{}
\bar{\Theta}=2.945 \times 10^{-6}\left(\frac{Z}{100}\right)^{2}
\left(\frac{10\lambdabar_{e}}{\Lambda}\right)^{3}
\sqrt{\left(\frac{300\text{K}}{T}\right){}
\left(\frac{P}{1\text{bar}}\right){}
\left(\frac{L}{30\text{m}}\right)} \,\,\,\,({\rm rad})\,.
\eear{}
It shows  the RMS angle is about 
$3$ $\mu$rad at  $Z=100, T=300\text{K}, 
P=1\text{bar}$, and $L=30\text{m}$, with 
 the cutoff at $\Lambda=10\lambdabar_{e}$. Note 
that this
value is 2 orders of magnitude larger than the bending angle by a 
single nucleus of the same $Z$ value and at the  impact parameter 
$b=\Lambda$.
Obviously, a greater amplification  can be obtained with a colder,
higher pressure  gas in a longer chamber. 
This demonstrates that a Coulomb gas can be an amplifier for 
the light bending in a Coulomb background.  

Let us now focus on the cutoff $\Lambda$. For a photon  at
the impact parameter $b$ with a nucleus
there are two constraints on the photon wavelength $\lambda${}, 
arising from the Euler-Heisenberg 
Lagrangian (\ref{lagrangian}) and the bending angle (\ref{bendingangle}). 
For the local Euler-Heisenberg interaction to be valid the wavelength should 
 be larger than the electron Compton length $\lambdabar_{e}$, and for the
  bending angle (\ref{bendingangle}) be valid the wavelength should be
  smaller than the impact parameter $b$, so that  the geometrical optics will
  be applicable. Thus
  \bear{}
  \lambdabar_{e} \ll \lambda \ll b\,,
  \eear{}
 which indicates the  cutoff  $\Lambda$, the lower bound of $b$,
  should be a multiple of the 
 photon wavelength. Putting in Eq. (\ref{result})
 \bear{}
 \Lambda =\mu\lambda{}\,,
 \eear{}
 where $\mu$ is a constant larger than unity,
 we get the RMS angle for a photon 
 beam of wavelength $\lambda${}:
 \bear{}
\bar{\Theta}=\frac{
Z^{2}\alpha^3}{160\mu^{3}}\sqrt{\frac{130\pi}{3}}
\sqrt{\frac{L}{L_{0}(P,T,\lambda)}}
\left(\frac{\lambdabar_{e}}{\lambda}\right)^{4}\,,{}
\label{result-1}
\eear{}
where $L_{0}(P,T,\lambda)=k_{B}T/P\lambda^{2}$, and
\bear{}
\bar{\Theta}=2.945 \times 10^{-6}\left(\frac{Z}{100}\right)^{2}
\left(\frac{10\lambdabar_{e}}{\mu\lambda}\right)^{3}
\sqrt{\left(\frac{300\text{K}}{T}\right){}
\left(\frac{P}{1\text{bar}}\right){}
\left(\frac{L}{30\text{m}}\right)} \,\,\,\,({\rm rad})\,.
\label{rms2}
\eear{}

The magnitude of the RMS angle has a sharp
 dependence on the cutoff parameter $\mu$.{}
This clearly results from the limitation of
the bending angle (\ref{bendingangle}) which is
 obtained in the geometrical 
optics and is valid only for $b\gg \lambda$.{}
At $b$ not larger than the wavelength, a more complete formula for
the bending angle may be obtained
using the wave optics, which 
is beyond the scope of this paper.
Physically, however, it is clear that the quartic divergence 
at small $b$, which gives the strong $\mu$ dependence, should disappear 
in a complete formula,
because at $b=0$ the bending angle must vanish for symmetry reasons.
We thus expect the bending angle will have a maximum at the impact parameter
$b=\mu_{0}\lambda$, where $\mu_{0}$ is a constant. We may then identify
$\mu_{0}$ with the cutoff parameter $\mu$.{}
It may not be  unreasonable to assume $\mu$ to be of the order of unity; so if we put
$\mu=5$ and $\lambda=5\lambdabar_e$, as an example, 
the RMS angle is a few tenths of $\mu$rad{}
at the reference values of $Z,P,T$ and $L$ in Eq. (\ref{rms2}).

An experimental consequence of the bending
will be a broadening of the beam as it travels through the gas. 
This beam broadening may be exploited to detect the light bending,
 by studying its dependence
  on the temperature, the pressure, the beam's travel distance,  and the
wavelength of the photon in the beam, and
see if it follows Eq. (\ref{result-1}). 
Another interesting signal would be the intensity profile of the
beam cross section.
Because of the random nature of the bendings off the nuclei the
intensity profile should be Gaussian.

To conclude, we have shown that the
 Coulombic light bending can be amplified 
by orders of magnitude with a photon beam 
traveling in a Coulomb gas. The RMS
bending angle at a given temperature and pressure is proportional to the
square root of the distance the beam traveled. 
Although we considered a Coulomb gas as the amplifying medium for the light
bending, it is conceivable that a similar amplification would occur as well
in other mediums, like  solid metals of large atomic number such as gold. 
Such an amplification might be investigated  in  experiments for 
precision measurement of refractive index
with $\gamma$ ray \cite{koga,kawasaki,gunther}, as  
the Coulombic bending can have a dominating effect on
the Delbr{\"u}ck scattering ~\cite{lee3}.

\begin{acknowledgments}
I am thankful to S. Han for encouragement. 
The early stages of this work were done in collaboration with 
N. Ahmadiniaz. 
This research was supported by the Basic Science 
Research Program through the National 
Research Foundation of Korea
(NRF), funded by the Ministry of 
Education, Science, and Technology
(2012R1A1A2044543).
\end{acknowledgments}

\bibliographystyle{apsrev4-1}
\bibliography{lightbending-coulombgas}

\end{document}